# Silabenzene Incorporated Covalent Organic Frameworks

Kewei Sun[1], Orlando J. Silveira[2], Yujing Ma[1], Yuri Hasegawa[3], Michio Matsumoto[4], Satoshi Kera[3], Ondřej Krejčí[2], Adam S. Foster*[2,5], Shigeki Kawai*[1,6]

[1]*Research Center for Advanced Measurement and Characterization, National Institute for Materials Science, 1-2-1 Sengen, Tsukuba, Ibaraki 305-0047, Japan*

[2]*Department of Applied Physics, Aalto University, Finland*

[3]*Institute for Molecular Science, Department of Photo-Molecular Science, Myodaiji, Okazaki 444-8585, Japan*

[4]*International Center for Materials Nanoarchitectonics (WPI-MANA), National Institute for Materials Science, 1-1, Namiki, Tsukuba, Ibaraki 305-0044, Japan.*

[5]*WPI Nano Life Science Institute (WPI-NanoLSI), Kanazawa University, Kakuma- machi, Kanazawa 920-1192, Japan*

[6]*Graduate School of Pure and Applied Sciences, University of Tsukuba, Tsukuba 305-8571, Japan*

## Abstract

Covalent organic frameworks (COFs) are a promising material for various applications such as gas storage/separation, catalysis, and energy storage, besides offering a confined space for chemical reaction. The introduction of unconventional elements into their frame structures and expanding their structural scope remains a major challenge in COF chemistry. Here, we present syntheses of two-dimensional and liner COFs substructures linked with 1,4-disilabenzene ($C_4Si_2$) by co-depositing silicon atoms and bromo-substituted poly aromatic hydrocarbons on Au(111). A combination of high-resolution scanning tunneling microscopy, photoelectron spectroscopy and density functional theory calculations reveal the detailed structures of the Si-incorporated COF, as well as its chemical properties. We find that each Si in a hexagonal $C_4Si_2$ ring is terminated by one Br atom. Furthermore, the $C_4Si_2$ ring can be transformed to the $C_4Si$ pentagonal ring by annealing. Silabenzene incorporated COFs may open up new possibilities in organosilicon chemistry and a wide range of practical applications.

## Introduction

Covalent organic frameworks (COFs), as a large class of porous organic materials have attracted intense research in past few decades due to the great potential for applications in the fields of, for example, gas storage and separation (*1,2*), catalysis (*3,4*) and optoelectronics (*5,6*). Since the seminal work of in-solution COF synthesis by Yaghi and his co-workers in 2005 (*7*), various kinds of COFs, composed of light elements (B, C, N, O, H), have been synthesized by Schiff base reactions, self-condensation of boronic acid, as well as coupling between boronic acid and catechol (*1, 7-10*). If the heavy elements are substituted in the precursor molecules, the elemental variety drastically enhances (*11*).

Silicon is an element of group 14 in the periodic table. Its four valence electrons at the outermost shell give it similar properties to carbon, yet the longer bond length and possible higher bonding states leads to unique chemical properties (*12*). Silicon incorporated organic functional molecules and novel nanostructures have attracted much attention in recent decades. For instance, silicates (Si-O) are some of the most studied compounds (*13,14*) and the successful synthesis of the silicate-COF was recently demonstrated (*15*). Silyl group commonly used as a protecting group in solution (*16*) was found to induce coupling reactions on surface (*17,18*). In contrast, silabenzene having unique heterocyclic rings with C-Si bonds (*19*) is still studied as a challenging target for the organic synthesis due to their high reactivities (*19-21*). Therefore, fabrication of two-dimensional COFs based on silabenzene has yet to be achieved. Here, we present silabenzene-incorporated COFs by linking brominated poly aromatic hydrocarbons (PAHs) and silicon atoms on Au(111). Their structures and chemical properties are analyzed by a combination of bond-resolved scanning tunneling microscopy (STM), scanning tunneling spectroscopy (STS), photoelectron spectroscopy and density functional theory (DFT) calculations. The synthesis of the COFs by on-surface coupling of Si atoms and PAHs may pave a way for fabrication of novel low-dimensional nanostructures.

## Results and Discussion

We employed on-surface synthesis to realize the silabenzene-incorporated COF under ultra-high vacuum conditions (*22,23*). This strategy has proven to be an important bottom-up method in the successful synthesis of graphene nanoribbons (GNR) with different edges, including different atomic species (*24,25*). Once GNRs are fused with each other at their edges, two dimensional

COFs can be synthesized (*26,27*). Alternatively, COFs have also been synthesized with predefined and small precursors (*28,29*). However, the structure of the precursor plays such a decisive role in this bottom-up approach that when no suitable precursor molecules are available, as is the case for unstable silabenzene, it becomes impossible to proceed.

Here, we overcome this limitation by combining conventional surface science techniques and on-surface chemistry. We employed 2,3,6,7,10,11-hexabromotriphenylene (HBTP) as a building block to fabricate two-dimensional Si-incorporated COFs (Fig. 1A). Firstly, a submonolayer $Au_xSi$ film was formed on Au(111) by depositing Si atoms at room temperature with a post-anneal to 420 K (Fig. S1). HBTP molecules were deposited on the substrate held at 420K, causing debromination. Consequently, the sample was further annealed at a higher temperature of 580 K. We found the formation of hexagonal porous structures in the STM topography (Fig. 1B). The pores surrounded by six bright spots are separated by 1.75 ± 0.02 nm (Fig. 1C). The contrast of the nodal site and surrounding six bright spots changes with respect to the bias voltage (Fig. S2). Since following a similar growth procedure method, but without pre-deposited Si atoms on a clean Au(111) resulted in formation of disordered films (Fig. S3), we conclude that Si atoms play a decisive role in the synthesis of the porous nanostructure. Before the final step of annealing at 580 K, we also found the formation of $SiBr_x$ (x = 1, 2, 3) compounds on the Au terrace as well as on the porous structure (Fig. S4) – these can be desorbed as $SiBr_4$ molecules from the surface by annealing at 450 K (*30*). In order to resolve the inner structures of the porous nanostructure, the tip apex was terminated by a CO molecule (*31,32*). The bond-resolved STM image taken at a constant height mode indicates that the porous structure was composed of six triphenylene backbones at the nodal site (Fig. 1D and Fig. S5), which are inter-connected by two different types of bonds. The length of the longer one (310 ± 20 pm) is far more than the typical length of a covalent bond, indicating it is composed of more than two atoms. We tentatively assigned this longer line as C-Si-X (X= Br or H) bonds, where Br atoms are from HBTP molecules, while H atoms are possibly from the chamber environment. In our previous study, Si and Br atoms can easily form a covalent bond on Au(111) (Fig. 1A) (*30*). The central bond between two neighbouring triphenylene backbones has a shorter length (280 ± 20 pm), and we suggest that this is due to deflection of the CO tip during scanning and is not a physical bond (*33*). In order to investigate the structure behind these images, we undertook an extensive DFT analysis of possible molecular assemblies, including a wide variety of configurations and atoms in the network. The best agreement is shown in Fig. 1E, suggesting the Si atoms in the C$_4$Si$_2$ ring are passivated by Br atoms (Si atoms at the edge of porous structure can also bond to two Br atoms (Fig. S6)). We see a very good agreement with experiment in the STM simulated topography (Fig. 1F), with pores also separated by 1.75 nm. We also see good agreement in the comparison between the high resolution, CO tip STM images – the simulated image in Fig. 1G reproduces the sharp contrast over the central triphenylene backbone, shows a bond between neighbouring triphenylenes (293 pm) and also a long bond over C-Si-Br (333 pm). Overall, this confirms that the triphenylene blocks are connected via the planar C$_4$Si$_2$ ring resulting in 1,4-disilabenzene-linked COFs (Si-COF).

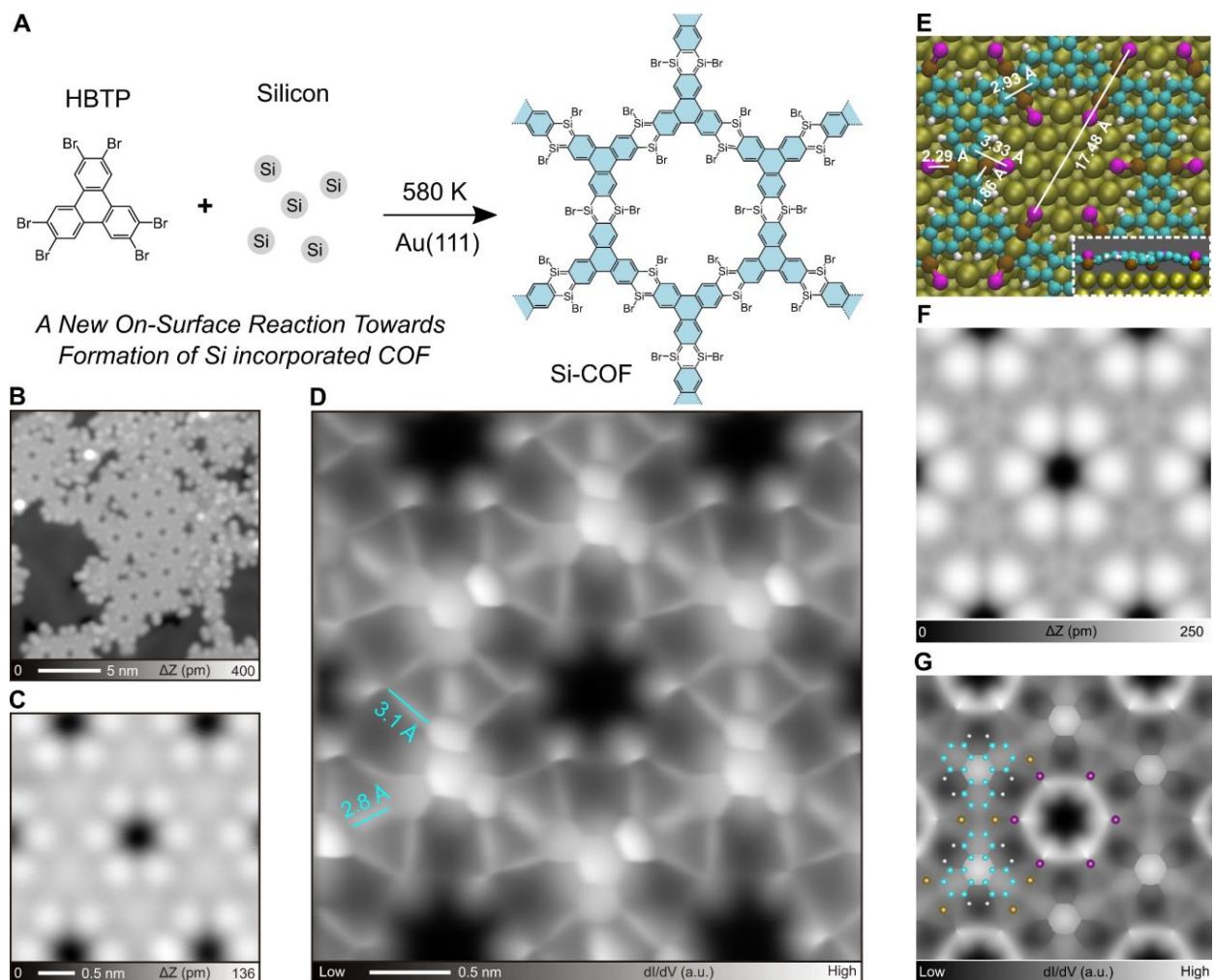

**Fig. 1. Synthesis of porous Si incorporated COF.** (**A**) Scheme of on-surface reaction for aryl-Si coupling reaction on Au(111). (**B**) STM topography after annealing at 580 K. Sample bias voltage $V = 100$ mV and tunneling current $I = 20$ pA. (**C**) A close-up STM image of Si-COF. $V = 200$ mV and $I = 5$ pA. (**D**) high-resolution constant height d$I$/d$V$ map. (**E**) Predicted structure (side view inset) from DFT simulations with key bond lengths (Å) indicated. Atom colours are carbon (cyan),

hydrogen (white), gold (yellow), silicon (brown) and bromine (purple). Associated simulated STM images showing **(F)** STM topography at 200 mV and constant charge density of $1\times10^{-8}$ e/Å$^3$ and **(G)** high-resolution image at a bias of 0.5 V and a height of 0.3 nm.

The electronic properties of Si-COF were measured in more detail by scanning tunneling spectroscopy (STS) (Fig. 2A). We found occupied and empty states around -670 mV and +1.4 V, respectively, resulting in a band gap of $2.1 \pm 0.1$ eV. Si-Br assigned spots are brighter than triphenylene blocks at occupied state (Fig. 2B), while the opposite is observed at unoccupied state (Fig. 2C). The electronic states of Si-COF on Au(111) have been detected by STS characterization (Fig. S7). The calculated density of states (Fig. 2D) shows similar features to the STS, with matching rapidly rising DOS around -1.5 and 2.0 V. This rapid rise is associated with the states coming from the molecular backbone of the COF, as the contribution from the C atoms significantly increases, while the contributions from the Si and Br atoms remain more or less constant in the interval of energy shown in Fig. 2D. The direct comparison of the simulated occupied states image at -0.3 V (Fig. 2E) reproduces the experimental dominance of the Br atoms, with nothing seen on the molecular backbone. Fig. S8 in the SI shows that the Br dominance is actually observed in several occupied states within the window of -1.0 and 0.0 V, and in the empty states as well, up to 0.8 V. For higher biases, however, contributions from both Br atoms and molecular backbone are observed uniformly, until the general feature previously observed is inverted. Fig. 2F, for example, shows this inversion at 2.2 V, where the strong dominance of the molecular backbone seen in the experiment is finally reproduced.

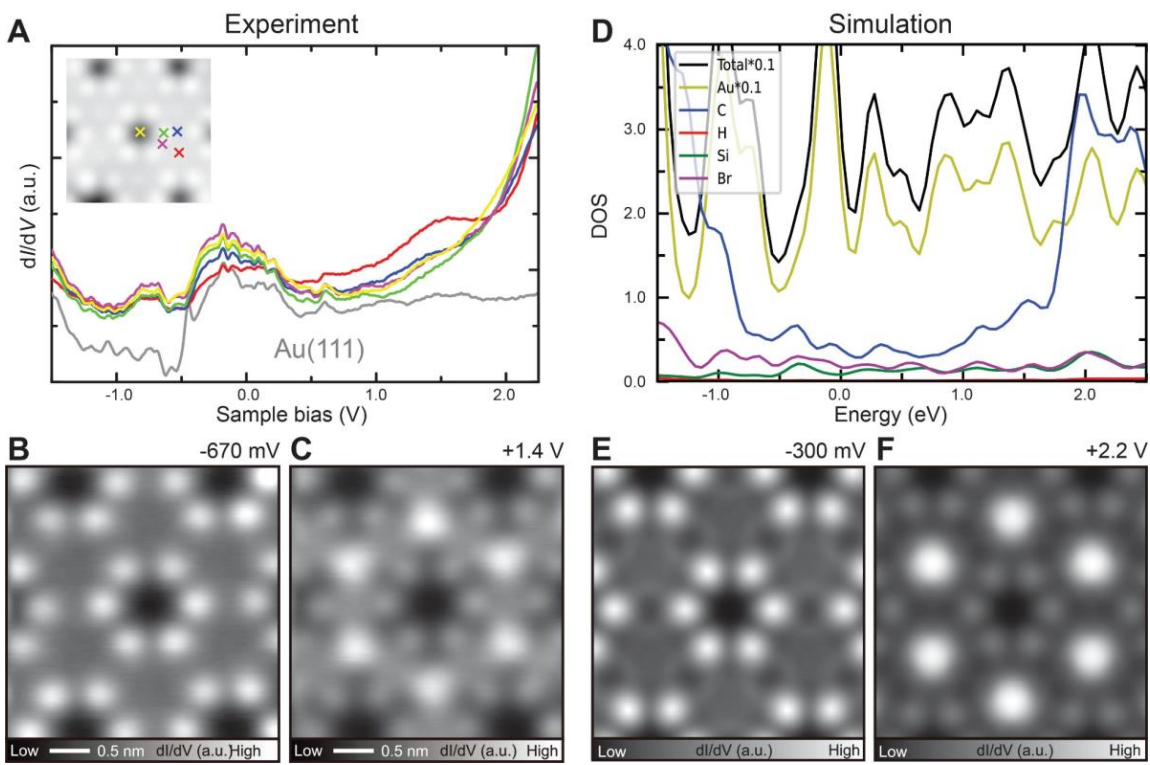

**Fig. 2**. (**A**) d$I$/d$V$ curves were recorded above the Si-COF (inset image) and Au (111) surface for a contrast. Constant height d$I$/d$V$ maps were measured with a CO terminated tip at (**B**) -670 mV and (**C**) +1.4 V. (**D**) Calculated Density of States (DOS) and simulated constant height d$I$/d$V$ maps at a bias of (E) -0.3 V and (F) +2.2 V.

In order to investigate the chemical properties of the Si-COF, we carried out synchrotron photoemission spectroscopic measurements after each reaction step. The clean surface of Au(111) substrate was first ensured by the presence of well-defined spin-orbit doublet peaks of Au $4f_{7/2}$ (83.8eV) and Au $4f_{5/2}$ separated by 3.69eV, which is in excellent agreement with established numbers (Fig. 3A).(*34-36*). After depositing Si atoms on a clean Au(111) surface, Au 4f doublet peaks(Fig. 3A) were significantly broadened by other components with a higher BE of 0.51 eV, which corresponds to the AuSi$_x$ alloy (*37-39*) (Fig. S9). After the synthesis of the COFs, the Au 4f spectra became comparable to that of clean Au 4f doublet peaks (Fig.3A). As observed in the STM measurements, the dissociated Br atoms from HBTP molecules tend to react with Si atoms and consequently remove the AuSi$_X$ layer from the surface by forming highly-volatile SiBr$_4$ molecules (*30*). This phenomenon was also identified in the Br 3d spectra (Fig. S11) where we observed a significant reduction in the signals of Br 3d after further annealing at higher temperature.

The corresponding Si 2p spectra of the AuSi$_x$ layer on Au(111) shows that the characteristic Si 2p doublet peaks with a small separation from p-orbitals spin-orbit coupling (0.61 eV) as the Si 2p$_{3/2}$ peak locates at 99.75 eV (Fig.3B, Fig. S11). We assume that charge transfer from Si to Au is responsible for the small shift to higher energy, compared with a value measured in the bulk (99.3 eV) (*40*). After the synthesis of the COFs, the Si 2p spectrum became complex. Our best fitting is consistent with four sets of doublet peak components, which can be associated to three different charge states of the Si atom (in our case, those components locate 0.71 eV, 2.0 eV and 3.18 eV higher in BE than elemental Si respectively). Although these numbers differ slightly from those of integer charge states of Si$^{1+}$(1 eV), Si$^{2+}$(1.81 eV), and Si$^{3+}$(2.63 eV) measured in bulk inorganic form (*41*,*42*), the Si-COF exhibits two major species with relatively large peak area that locates 0.71 eV and 2.0 eV higher BE than the elemental Si position. Since the component shifted by 2 eV is almost comparable to a Si$^{2+}$ charge state (1.81eV), it is reasonable to assign it with the structure proposed in illustration (Fig.1A) where the Si atoms covalently bonded with two carbon atoms and as well as a bromine atom as indicated from the bond-resolved STM image (Fig. 1D) and theoretical calculations (Fig. 1E), as both carbon and bromine have a higher electronegativity than Si, therefore a slightly higher charge state than Si$^{2+}$, but less than Si$^{3+}$, can be expected for these C$_4$Si$_2$Br$_2$ linker units. For the other major component that shifted by 0.71 eV in BE, we attributed it to the Si atoms that are incorporated to the edges of the Si-COF, bonded with one carbon and one Br atoms (Fig. S6). For the peak at 3.18 eV, it should be still from the Si at the edges, which bonded with one carbon and two Br atoms (Fig. S6).

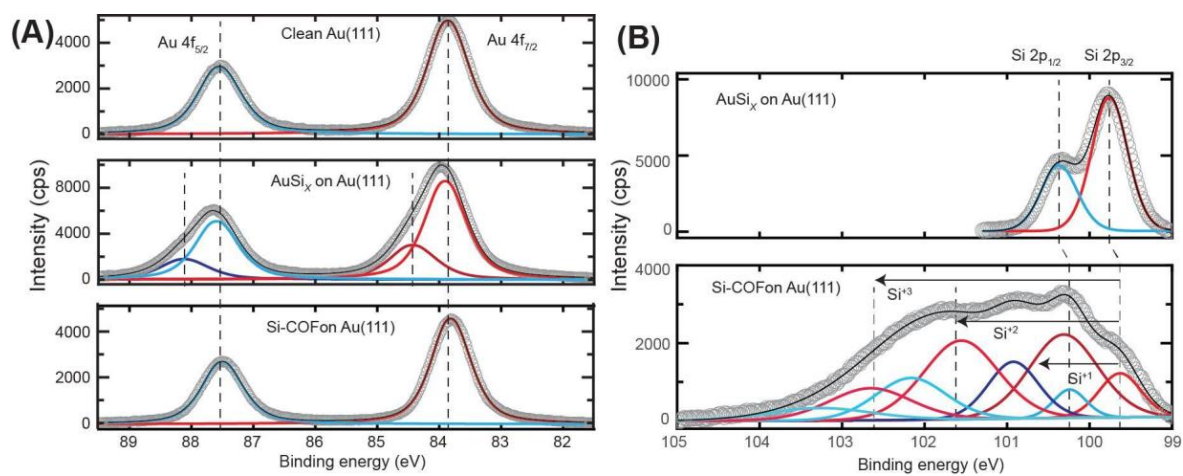

**Fig.3. Photoemission spectroscopy measurement.** (**A**) Au 4f and (**B**) Si 2p core level spectra at each reaction step.

In order to demonstrate a generality of the Si-C bond formation by co-deposition of Si atoms and bromo-substituted molecules, we employed 4,5,9,10-tetrabromopyrene (TBP) molecules, in which two groups of ortho-bromine atoms are introduced at both sides of pyrene backbone (Fig. 4A). After TBP molecules were deposited on the partially-covered $AuSi_x$ layer on Au(111) held at 420 K, short oligomers appeared (Fig. 4B). The close-up view (Fig. 4C) shows that bright dots are located at the edges of the longitudinal axis. A constant height d$I$/d$V$ image taken with a CO-terminated tip (Fig. 4D) and the corresponding Laplace filtered image for enhancement of bond features (Fig. 4E) reveal the detailed structures, in which the pyrene backbones are connected via disilabenzene. The bright dots indicated by arrows in Fig. 4E correspond to Br atoms. Comparison to the simulated structure and STM images (Fig. 4F) support that a Si-doped cove-edge graphene nanoribbon (Si-cove GNRs) was synthesized. Note that the brighter dots indicated by circles in Fig. 4B can be assigned to the second Br atoms bonded with Si after C-Si-C coupling (Fig. S12), which also can be removed by tip manipulation (Fig. S13).

After annealing at a higher temperature of 580 K, the structure of oligomers further changed as the bright dots at the edges form a zigzag arrangement along the longitudinal axis (Fig. 4G,H). The high-resolution constant height d$I$/d$V$ image (Fig. 4I) and the corresponding Laplace filtered image (Fig. 4J) show that the $C_4Si_2$ six-membered rings were transformed into the $C_4Si$ five-membered rings. Unlike the $C_4Si_2$ rings in Si-COF, the $C_4Si_2$ six-membered rings in Si-cove GNRs are not stabilized within the network structure of COF, and thus sequential cyclization from desilicification of the disilabenzene and subsequent dehydrogenation of the pyrene backbones proceed upon thermal activations. Again, the simulated structure and STM images support our analysis (Fig. 4K). Hence, the high reproducibility of the Si-incorporated COF structure is unambiguously proven.

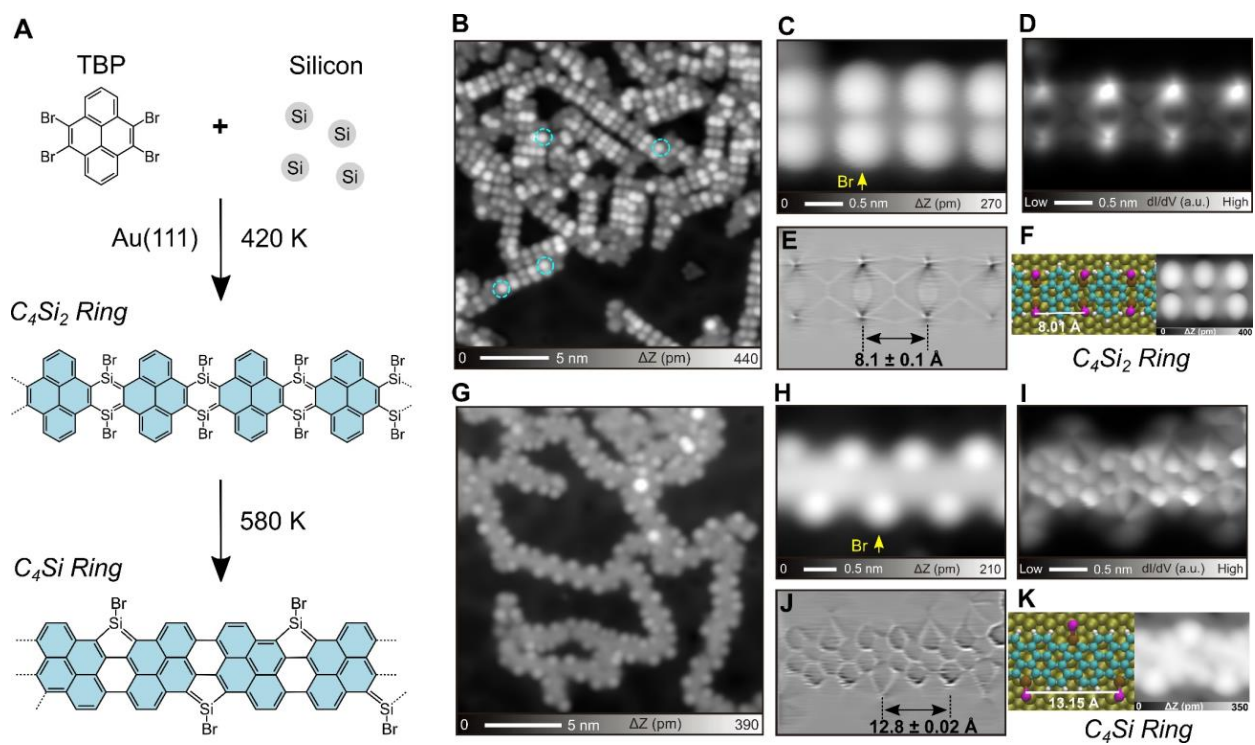

**Fig. 4.** (**A**) The scheme of on-surface synthesis of two types of Si-doped GNRs on Au(111). (**B**) STM topography of Si-covered Au(111) held at 420 K after depositing TBP molecules. Sample bias voltage $V = 200$ mV and tunneling current $I = 5$ pA. (**C**) A close-up view of Si-Cove GNR. $V = 200$ mV and $I = 5$ pA. (**D**) Constant height d$I$/d$V$ image of the area in (c), and (**E**) the corresponding Laplace filtered image. (**F**) The simulated chemical structure of Si-Cove GNR and associated simulated STM image at 200 mV and constant charge density of $1\times10^{-8}$ e/Å$^3$. (**G**) STM topography of sample after annealing at 580 K. $V = 100$ mV and $I = 10$ pA. (**H**) A closeup view of Si-Armchair GNR. $V = 100$ mV and $I = 20$ pA. (**I**) Constant height d$I$/d$V$ image of the area in (H), and (**J**) the corresponding Laplace filtered image. (**K**) The simulated chemical structure of Si-Armchair GNR and associated simulated STM image at 200 mV and constant charge density of $5\times10^{-8}$ e/Å$^3$.

**Discussion**

In summary, we synthesize silabenzene-incorporated COFs by reacting bromo-substituted molecules HBTP and Si atoms on Au(111). The linked structure of C$_4$Si$_2$ rings passivated by Br atoms can be determined by bond-resolved STM images combined with DFT calculations as well as XPS measurements. The TBP molecules also can form the $C_4Si_2$ ring after reacted with Si atoms, which even was transformed into the $C_4Si$ ring after desilicification and dehydrogenation. These results demonstrate a high-generality of the C-Si on-surface coupling by co-depositing Si atoms and PAHs on Au(111), which may further extend syntheses of various low-dimensional nano nanostructures.

**Supporting Information**

**AUTHOR INFORMATION**

**Corresponding Author**

*adam.foster@aalto.fi

* KAWAI.Shigeki@nims.go.jp

**Notes**

The authors declare no competing financial interest.

**ACKNOWLEDGEMENTS**

This work was supported in part by Japan Society for the Promotion of Science (JSPS) KAKENHI Grant Number 19H00856, 21K18885, and 21F21058. The authors thank staff members of the UVSOR Synchrotron Facility for the support during the XPS measurements. O. K. was supported by the European Union's Horizon 2020 research and innovation programme under the Marie Skłodowska-Curie grant agreement No. 845060

**Author contributions:** S. Kawai conceived the project. K. Sun performed the STM measurements and analyzed the data. Y.J., S. Kera, and S. Kawai conducted the XPS experiments and Y.M. and

S. Kera analyzed the data. M.M. synthesized the precursor molecules. O. J. S., O.K., and A.S.F. conducted theoretical calculations. K. Sun, O.J.S., A.S.F. and S. Kawai contributed to writing the manuscript. All authors commented on the manuscript.

**Competinginterests:** The authors declare that they have no competing interests.

**Data and materials availability**: All data needed to evaluate the conclusions in the paper are present in the paper and/or the Supplementary Materials. Additional data related to this paper may be requested from the authors.

**References**
1. Furukawa, H.; Yaghi, O. M. *J. Am. Chem. Soc.* **2009**, *131*, 8875–8883.
2. Kang, Z.; Peng, Y.; Qian, Y.; Yuan, D.; Addicoat, M. A.; Heine, T.; Hu, Z.; Tee, L.; Guo, Z.; Zhao, D. *Chem. Mater.* **2016**, *28*, 1277−1285.
3. Ding, S.; Gao, J.; Wang, Q.; Zhang, Y.; Song, W.; Su, C.; Wang, W. *J. Am. Chem. Soc.* **2011**, *133*, 19816–19822.
4. Lin, S.; Diercks, C. S.; Zhang, Y.; Kornienko, N.; Nichols, E. M.; Zhao, Y.; Paris, A. R.; Kim, D,; Yang, P.; Yaghi, O. M.; Chang, C. J. *Science* **2015**, *349,* 1208−1213.
5. Spitler, E. L.; Dichtel, W. R. *Nat. Chem.* **2010**, *2*, 672−677.
6. Wan, S.; Guo, J.; Kim, J.; Ihee, H.; Jiang, D. *Angew. Chem. Int. Ed.* **2008**, *47*, 8826–8830.
7. Côté, A. P.; Benin, A. I.; Ockwig, N. W.; O'Keeffe, M.; Matzger, A. J.; Yaghi, O. M. *Science* **2005**, *310*, 1166−1170.
8. Geng, K.; He, T.; Liu, R.; Dalapati, S.; Tan, K. T.; Li, Z.; Tao, S.; Gong, Y.; Jiang, Q.; Jiang, D. *Chem. Rev.* **2020**, *120*, 8814−8933.
9. Ding, S.; Wang, W. *Chem. Soc. Rev.* **2013**, *42*, 548−568.
10. Feng, X.; Ding, X.; Jiang, D. *Chem. Soc. Rev.* **2012**, *41*, 6010−6022.
11. Narita, A.; Wang, X.; Feng, X.; Müllen, K. *Chem. Soc. Rev.* **2015**, *44*, 6616−6643.
12. Silicon Chemistry: From the Atom to Extended Systems. Edited by Peter Jutzi and Ulrich Schubert. ISBN: 978-3-527-61076-1, May 2007.
13. Alexandre, M.; Dubois, P. *Mater. Sci. Eng. R Rep.* **2000**, *28*, 1–63.

14. Stein, A.; Melde, B. J.; Schroden, R. C. *Adv. Mater.* **2000**, *12*, 1403–1419.
15. Roeser, J.; Prill, D.; Bojdys, M. J.; Fayon, P.; Trewin, A.; Fitch, A. N.; Schmidt, M. U.; Thomas, A. *Nat. Chem.* **2017**, *9*, 977–982.
16. House, H. O.; Czuba, L. J.; Gall, M.; Olmstead, H. D. *J. Org. Chem.* **1969**, *34*, 2324–2336.
17. Liu, L.; Klaasen, H.; Witteler, M. C.; Lammers, B. S.; Timmer, A.; Kong, H.; Mönig, H.; Gao, H.; Neugebauer, J.; Fuchs, H.; Studer, A. *Nat. Chem.* **2021**, *13*, 350−357.
18. Sun, K.; Sagisaka, K.; Peng, L.; Watanabe, H.; Xu, F.; Pawlak, R.; Meyer, E.; Okuda, Y.; Orita, A.; Kawai, S. *Angew. Chem. Int. Ed.* **2021**, *60*, 19598–19603.
19. Tokitoh, N. *Acc. Chem. Res.* **2004**, *37*, 86-94.
20. Abersfelder, K.; White, A. J. P.; Rzepa, H. S.; Scheschkewitz, D. *Science*, **2010**, *327*, 564–566.
21. Chen, Y.; Li, J.; Zhao, Y.; Zhang, L.; Tan, G.; Zhu, H.; Roesky, H. W. *J. Am. Chem. Soc.* **2021**, *143*, 2212−2216.
22. Grill, L.; Hecht, S. *Nat. Chem.* **2020**, *12*, 115−130.
23. Grill, L.; Dyer, M.; Lafferentz, L.; Persson, M.; Peters, M. V.; Hecht, S. *Nat. Nanotechnol.* **2007**, *2*, 687−691.
24. Cai, J.; Ruffieux, P.; Jaafar, R.; Bieri, M.; Braun, T.; Blankenburg, S.; Muoth, M.; Seitsonen, A. P.; Saleh, M.; Feng, X.; Müllen, K.; Fasel, R. *Nature* **2010**, *466*, 470−473.
25. Ruffieux, P.; Wang, S.; Yang, B.; Sánchez-Sánchez, C.; Liu, J.; Dienel, T.; Talirz, L.; Shinde, P.; Pignedoli, C. A.; Passerone, D.; Dumslaff, T.; Feng, X.; Müllen, K.; Fasel, R. *Nature* **2016**, *531*, 489−492.
26. Moreno, C.; Vilas-Varela, M.; Kretz, B.; Garcia-Lekue, A.; Costache, M. V.; Paradinas, M.; Panighe, M.; Ceballos, G.; Valenzuela, S. O.; Peña, D.; Mugarza, A. *Science* **2018**, *360*, 199−203.
27. Kawai, S.; Saito, S.; Osumi, S.; Yamaguchi, S.; Foster, A. S.; Spijker, P.; Meyer, E. *Nat. Commun.* **2015**, *6*, 8098.
28. Zwaneveld, N. A. A.; Pawlak, R.; Abel, M.; Catalin, D.; Gigmes, D.; Bertin, D.; Porte, L. *J. Am. Chem. Soc.* **2008**, *130*, 6678−6679.
29. Galeotti, G.; De Marchi, F.; Hamzehpoor, E.; MacLean, O.; Rajeswara Rao, M.; Chen, Y.; Besteiro, L. V.; Dettmann, D.; Ferrari, L.; Frezza, F.; Sheverdyaeva, P. M.; Liu, R.; Kundu, A. K.; Moras, P.; Ebrahimi, M.; Gallagher, M. C.; Rosei, F.; Perepichka, D. F.; Contini, G. *Nat.*

*Mater.* **2020**, *19*, 874−880.

30. Sun, K.; Nishiuchi, T.; Sahara, K.; Kubo, T.; Foster, A. S.; Kawai, S. *J. Phys. Chem. C* **2020**, *124*, 19675−19680.
31. Gross, L.; Mohn, F.; Moll, N.; Liljeroth, P.; Meyer, G. *Science* **2009**, *325*, 1110−1114.
32. Temirov, R.; Soubatch, S.; Neucheva, O.; Lassise, A. C.; Tautz, F. S. *New J. Phys.* **2008**, *10*, 053012.
33. Neu, M.; Moll, N.; Gross, L.; Meyer, G.; Giessibl, F. J.; Repp, J. *Phy. Rev. B* **2014**, *89*, 205407.
34. Dückers, K. Bonzel, H. P. *Surf. Sci.* **1989**, *213*, 25−48.
35. Seah, M. P.; Gilmore, I. S.; Beamson, G. *Surf. Interface Anal.* **1998**, *26*, 642−649.
36. Turner, N. H.; Single, A. M. *Surf. Interface Anal.* **1990**, *15*, 215−222.
37. Sundaravel, B.; Sekar, K.; Kuri, G.; Satyam, P. V.; Dev, B. N.; Bera, S.; Narasimhan, S. V.; Chakraborty, P.; Caccavale, F. *Appl. Surf. Sci.* **1999**, *137*, 103–112.
38. Yeh, J. J.; Hwang, J.; Bertness, K.; Friedman, D. J.; Cao, R.; Lindau, I. *Phy. Rev. Lett.* **1993**, *70*, 3768–3771.
39. Molodtsov, S. L.; Laubschat, C.; Kaindl, G.; Shikin, A. M.; Adamchuk, V. K. *Phy. Rev. B* **1991**, *44*, 8850–8857.
40. Lu, Z. H.; Sham, T. K.; Norton, P. R. *Solid State Commun.* **1993**, *85*, 957–959.
41. Yeom, H. W.; Uhrberg, R. *Jpn. J. Appl. Phys.* **2000**, *39*, 4460–4463.
42. Hollinger, G.; Himpsel, F. J. *Phy. Rev. B* **1983**, *28*, 3651–3653.

# Supplementary Materials for
# Silabenzene Incorporated Covalent Organic Frameworks

Kewei Sun[1], Orlando J. Silveira[2], Yujing Ma[1], Yuri Hasegawa[3], Michio Matsumoto[4], Satoshi Kera[3], Ondřej Krejčí[2], Adam S. Foster*[2,5], Shigeki Kawai*[1,6]

[1]Research Center for Advanced Measurement and Characterization, National Institute for Materials Science, 1-2-1 Sengen, Tsukuba, Ibaraki 305-0047, Japan

[2]Department of Applied Physics, Aalto University, Finland

[3]Institute for Molecular Science, Department of Photo-Molecular Science, Myodaiji, Okazaki 444-8585, Japan

[4]International Center for Materials Nanoarchitectonics (WPI-MANA), National Institute for Materials Science, 1-1, Namiki, Tsukuba, Ibaraki 305-0044, Japan.

[5]WPI Nano Life Science Institute (WPI-NanoLSI), Kanazawa University, Kakuma- machi, Kanazawa 920-1192, Japan

[6]Graduate School of Pure and Applied Sciences, University of Tsukuba, Tsukuba 305-8571, Japan

Materials and Methods,
Supplementary Text
Fig. S1 to S13
Full Reference List

**Materials and Methods**

Experimental: Scanning Tunneling microscopic measurement

All the experiments were conducted in a low temperature scanning tunneling microscopy (STM) system (home-made) at 4.3 K under high-vacuum environment ($< 1 \times 10^{-10}$ mbar). The bias voltage was applied to the sample while the tip was electrically grounded. A Au(111) surface was cleaned through cyclic sputtering (Ar$^+$, 10 min) and annealing (720 K, 15 min). Si atoms were deposited on a clean Au(111) surface with an electron beam evaporator (SPECS GmbH). 2,3,6,7,10,11-hexabromotriphenylene (HBTP) was purchased from Sigma-Aldrich and further purified by recrystallization from *o*-dichlorobenzene. 4,5,9,10-tetrabromopyrene (TBP) was synthesized by following a literature procedure (*1*). HBTP and TBP molecules were deposited from Knudsen cells (Kentax GmbH). A STM tip was made from the chemically etched tungsten. For bond-resolved imaging, the tip apex was terminated by a small CO molecule picked up from the surface (*2*). The bias voltage was set close to zero voltage. The modulation amplitude was 7 mV$_{rms}$ and the frequency was 510 Hz.

Experimental: Photoemission spectroscopy measurement

Photoemission spectroscopy measurements were conducted at BL2B beamline in UVSOR-III Synchrotron, featuring a monochromatic light source with a photon energy ranging from 23 eV to 205 eV. The high-resolution Au 4f, Si 2p, and Br 3d spectra were taken with a photon energy of 130 eV, measured in normal emission with an energy resolution of below 65 meV. All spectra were processed with a Shirley background subtraction as the binding energies are respective to the Fermi edge of the Au(111) substrate. The core level spectra fit by GL(*m*) and SGL(*m*) functions represent the product and sum of Gaussian and Lorentzian functions, respectively. The parameter *m* indicates a ratio between the two functions as *m* = 0 is a pure Gaussian and *m* = 100 is pure Lorentzian.

Detailed fitting parameters of all XPS spectra are described in a table underneath each plot. Fittings followed general protocols, such as doublet peaks associated with various orbitals adopt well-defined energy separations. The Au 4f doublet components 4f$_{7/2}$ and 4f$_{5/2}$ have a large energy separation of 3.69 eV; Si 2p doublet components 2p$_{3/2}$ and 2p$_{1/2}$ are slightly separated by 0.61 eV in binding energy; and Br 3d doublet, 3d$_{5/2}$ and 3d$_{3/2}$ has an intermediate energy separation of 1.05 eV. The other restraints applied in spectra fitting are the area (intensity) ratios between the doublet components, where Au 4f doublet has an intensity ratio of 0.75 ($4f_{7/2} : 4f_{5/2} = 4:3$); Si 2p doublet has an intensity ratio of 0.5 ($2p_{3/2} : 2p_{1/2} = 2:1$); Br 3d doublet adopted the intensity ratio of 0.67 ($3d_{5/2} : 3d_{3/2} = 3:2$).

Our best fitting of the Au 4f spectra was well-reproduced with a line shape of SGL(55)T(7), which represent a Gaussian/Lorentzian sum form with a slight underlying asymmetric profile blend T(k). Si 2p doublet peaks are fitted with line shape of GL(30), it represents a Gaussian/Lorentzian product form. Detail mathematical equations are available in CasaXPS library. In each spectra plot, the raw XPS data and convoluted envelope obtained by fitting are indicated by hollow circle markers and semi-transparent green lines, respectively. Each individual component is indicated by different colored lines, where each doublet set is plotted with red and blue curves, and Shirley background is indicated with brown dash lines.

Si atoms were deposited on a clean Au(111) surface with an electron beam evaporator (SPECS GmbH). HBTP and TBP were deposited from a Knudsen cells (Kentax GmbH). The deposition and annealing parameters were the same as those in the STM measurement.

In Fig. S9b, upon Si deposition onto Au substrate, a discernable broadening of each Au 4f doublet peaks indicates the presence of two chemical components. The second set of doublet peaks were shifted by 0.51 eV to high binding energy compared to the bulk gold. Shifting to a higher BE for the gold-silicide ($AuSi_X$) alloy on the Au(111) was commonly reported. However, one might find it a bit contradictory that if charge flow from Si into Au as mentioned in the main text, yet both Au 4f and Si 2p spectra are shifting to the higher BE side. In fact, this phenomenon was well explained before and proposed with a d-electron depletion model (*3*). It is understood that on the account of Au-Si interaction, Si transfers s charge to Au, but Au loses more localized d-electrons forming a s-d hybrid bond, in which the key factor is that the Coulomb interaction between Au 4f and the 6s conduction electron is quite different from that of 4f and 5d interaction, with the latter one is actually significantly larger (by ~3eV) and affecting the overall binding states of Au.

In Fig. S10, the lower binding energy set (Br $3d_{5/2}$ locate at 68.76 eV) were assigned to disassociated Br atoms/clusters (*4*) and higher BE position ($3d_{5/2}$ locate at 69.71 eV) was in good agreement with previous studies (*5-7*). Compared to the C-Br bonding related Br

3d energy positions in these reports, our component at higher BE is highly accountable for Si-Br bond forming a silicon tetrabromide ($SiBr_4$) compound which can be easily desorbed with higher temperature annealing. In addition, it is worth mentioning that the bromine signal essentially vanished after annealing at high temperature of 580 K, which is possibly due to different experimental setups (e.g. higher annealing temperature) at beamline.

Theoretical calculations

All first-principles calculations on the gold substrate in this work were performed using the periodic plane-wave basis VASP code (*8,9*) implementing the spin-polarized Density Functional Theory. To accurately include van der Waals interactions in this system, we used the DFT-D3 method with Becke-Jonson damping (*10,11*) – various other van der Waals functionals were tested and no significant differences were observed. Projected augmented wave potentials were used to describe the core electrons (*12*) with a kinetic energy cutoff of 500 eV (with PREC = accurate). Systematic *k*-point convergence was checked for all systems with sampling chosen according to the system size. This approach converged the total energy of all the systems to the order of 1 meV. The properties of the bulk and surface of Au were carefully checked within this methodology, and excellent agreement was achieved with experiments. For calculations of the assemblies on the surface, a vacuum gap of at least 1.5 nm was used. A 3x3x1 *k*-point grid was used and the upper three layers of Au (five layers in total) and all atoms in the assemblies were allowed to relax to a force of less than 0.01 eV/Å. Atomic structure visualizations were made with the VMD package (*13*). Standard simulated STM images were calculated using the CRITIC2 package (*14,15*) based on the Tersoff−Hamann approximation (*16*). For the high-resolution CO tip STM images, we have made use of the FHI-AIMS code (*17*) with the previous optimized geometry as a single point calculation. For these calculations the Perdew–Burke–Ernzerhof exchange-correlation functional was used (*18*) with Γ k-point only and the standard "light" basis set. The high-resolution CO tip STM images were then computed by means of the PP-STM code with fixed tip, where the broadening parameter η was set to 0.2 eV (*19*). The CO tip was approximated by 13% of the signal coming from the *s* orbital and 87% originating from the $p_{xy}$ orbitals on the Probe Particle, which gave a good agreement with close-by CO-STM and CO-d*I*/d*V* images (*20*). Since for the further away CO-STM images also 50/50 $s/p_{xy}$ ratio was also reported (*21*), we show additional comparison of $s/p_{xy}$ ratios for all the calculated voltage d$I$/d$V$ images in Fig S.8. Density of states analysis was made using the VASPKIT package (*22*).

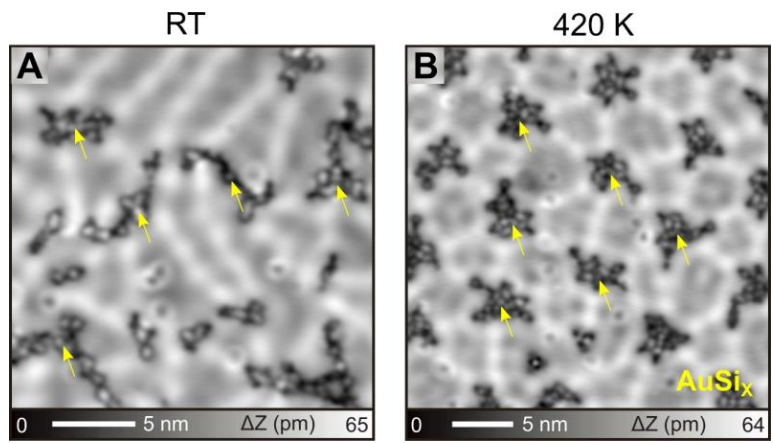

**Fig. S1.** (**A**) STM topography of Au(111) after depositing Si atoms kept at room temperature. The regions (indicated by arrows) look darker due to existence of Si atoms. (**B**) STM image of Au(111) covered by Si after annealing at 420 K. The darker regions whose sizes seem more homogeneous, are identified as AuSi$_x$ films on Au(111) surface, similar with previous report. Measurement parameters: Sample bias voltage $V$ = 100 mV and tunneling current $I$ = 100 pA in (**A**). $V$ = 100 mV and $I$ = 20 pA in (**B**).

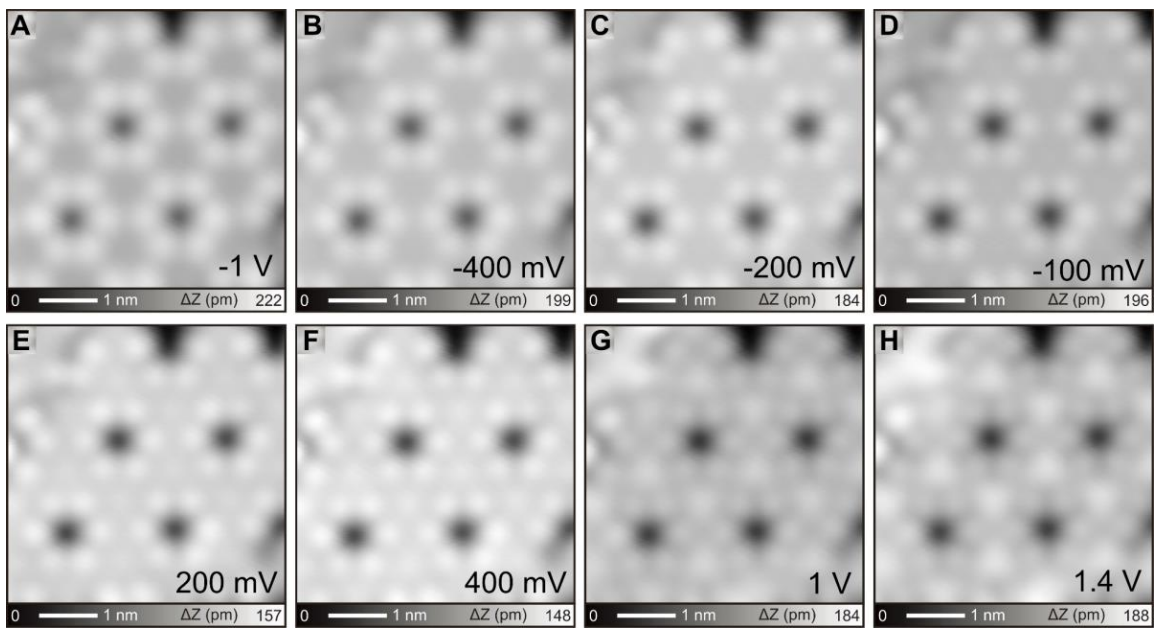

**Fig. S2.**(**A**)-(**H**) A series of STM topographies measured with different bias voltages.

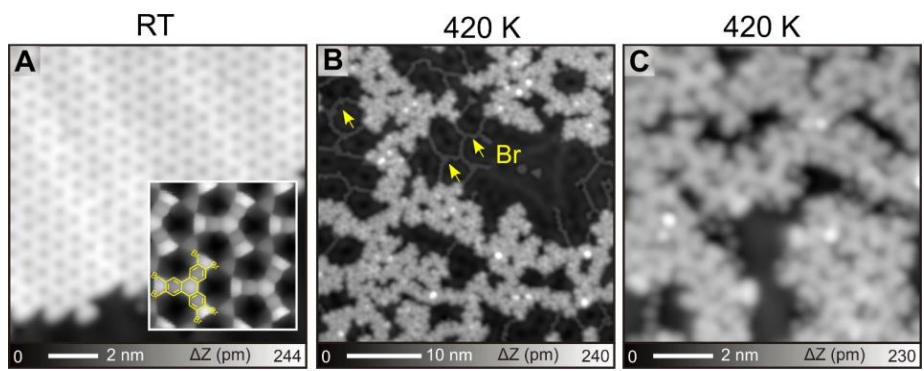

**Fig. S3.** (**A**) STM topography of Au(111) surface kept at room temperature after depositing HBTP molecules. Insert shows the bond-resolution constant height STM image of closeup view. (**B**) Typical STM image of Au(111) surface with HBTP molecules after annealing at 420 K for 10 minutes. Large scale irregular structures and scattered bromine atoms (indicated by arrows) after debromination can be observed. (**C**) A closeup view of irregular nanostructures after reaction of HBTP molecules. These results demonstrate that it is difficult to form well-ordered COFs by only HBTP molecules on Au(111). Measurement parameters: $V = 10$ mV and $I = 100$ pA in (**A**). $V = 200$ mV and $I = 20$ pA in (**B**). $V = 50$ mV and $I = 20$ pA in (**C**).

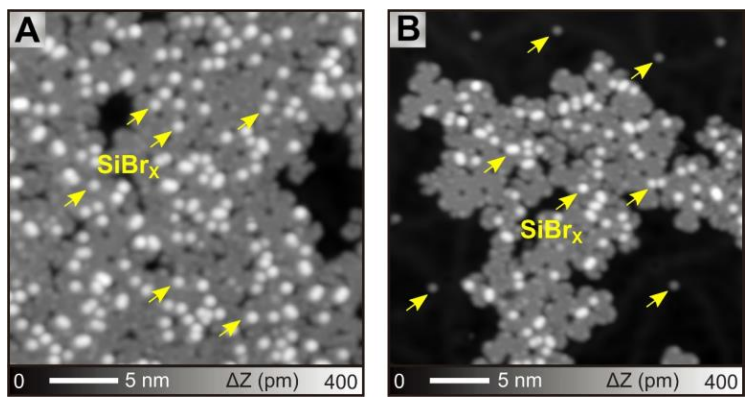

**Fig. S4.** STM topographies of porous structure on Au(111) after heating at 420 K. Many bright dots identified as SiBr$_X$ ($X = 1,2,3$) compounds on porous structure and Au(111) surface, indicated by arrows. Measurement parameters: $V = 200$ mV and $I = 10$ pA in (**A**). $V = 100$ mV and $I = 20$ pA in (**B**).

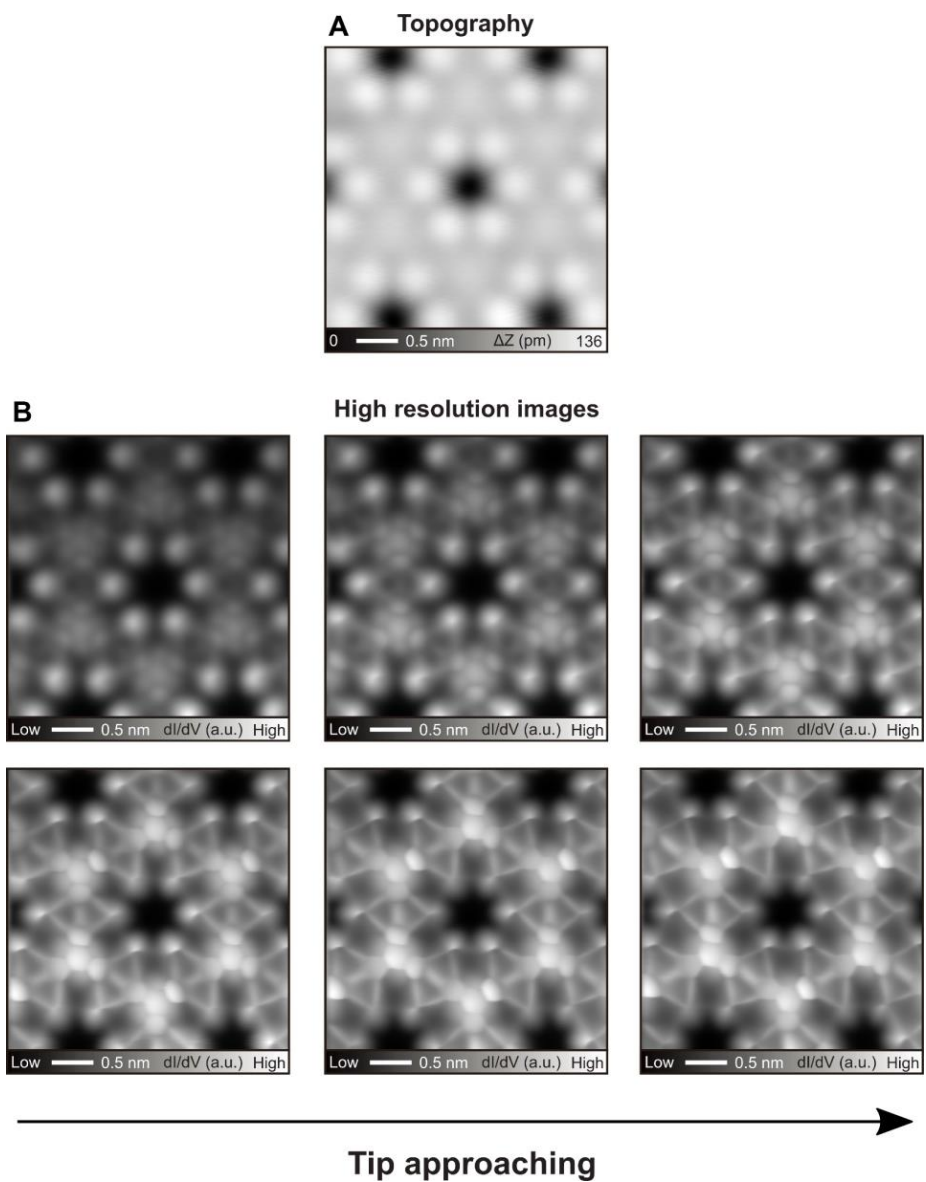

**Fig. S5.** (**A**) STM topography of the detailed Si-COF unit. (**B**) A series of high-resolution constant height d*I*/d*V* images with a CO-tip at different tip-sample distances. Measurement parameters: *V* = 200 mV and *I* = 5 pA in (**A**).

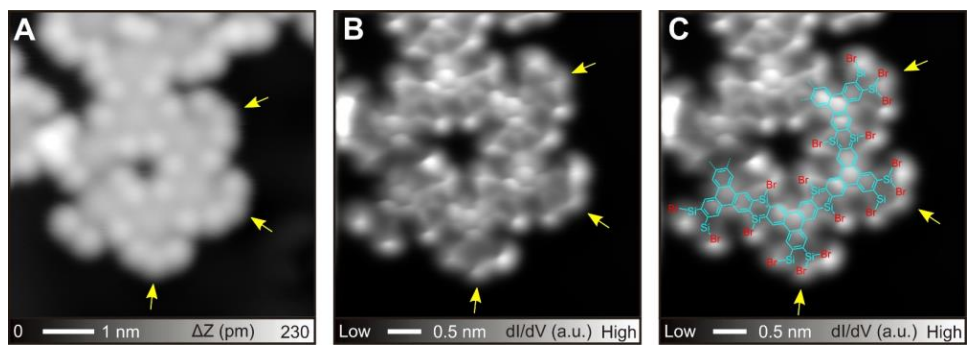

**Fig. S6.** (**A**) STM topography of single porous structure and (**B**) the corresponding high-resolution constant height d$I$/d$V$ image. (**C**) The chemical structure superimposed on high-resolution image. The arrows show the sites of additional Br atoms. Measurement parameters: $V = 200$ mV and $I = 5$ pA in (**A**).

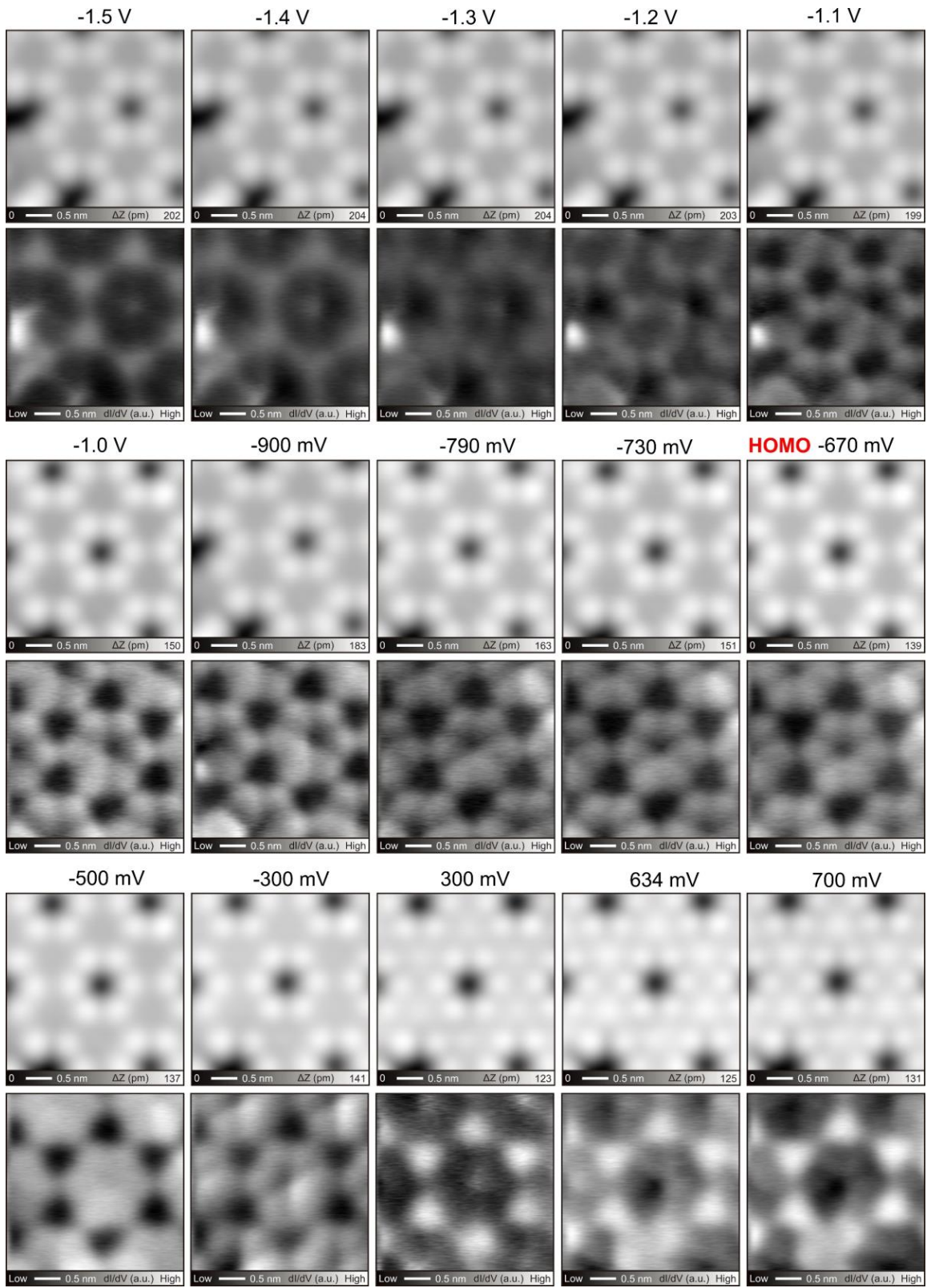

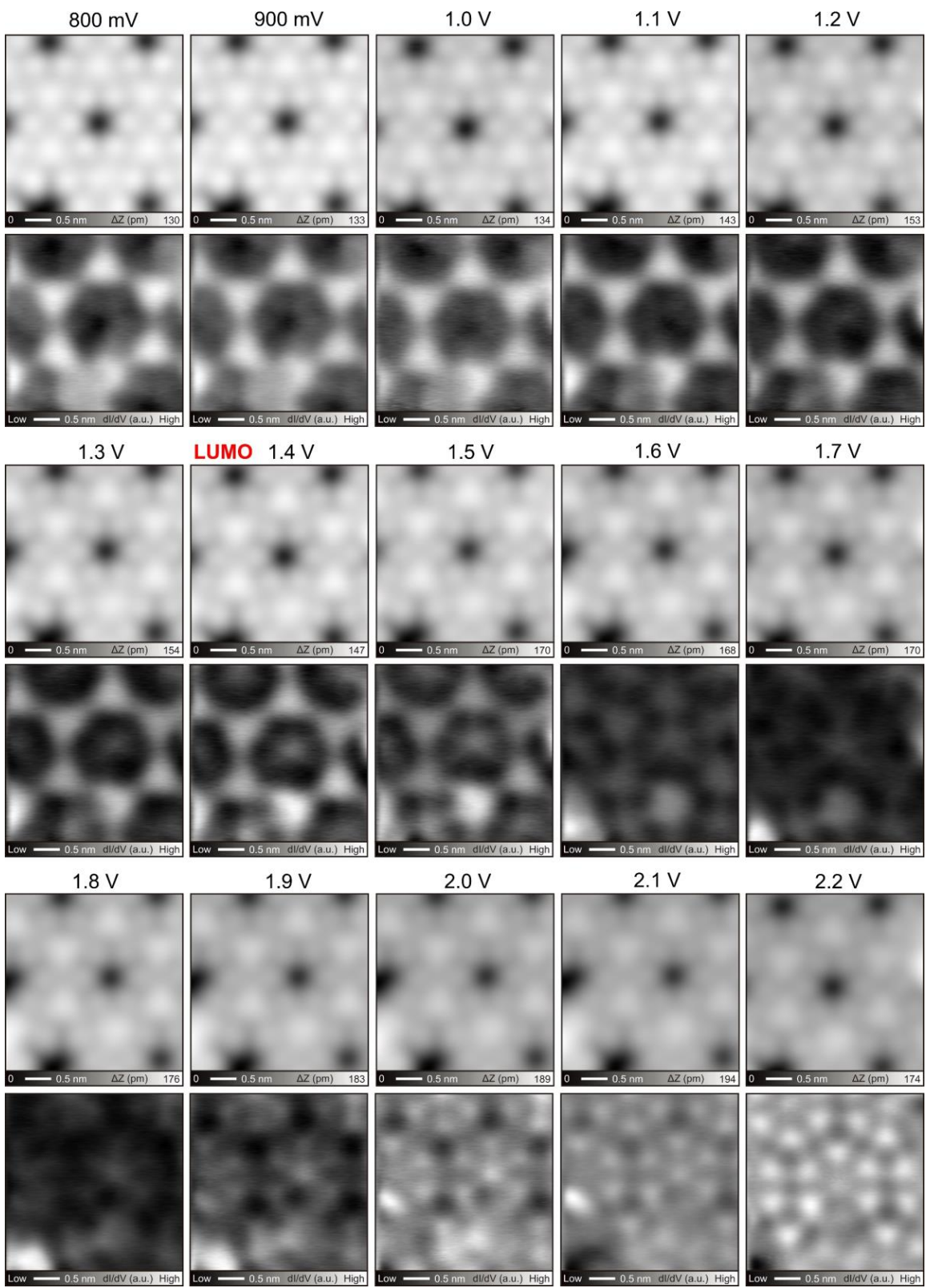

**Fig. S7.** A series of STM topographies of Si-COF and the corresponding constant current d*I*/d*V* maps measured at different bias voltages.

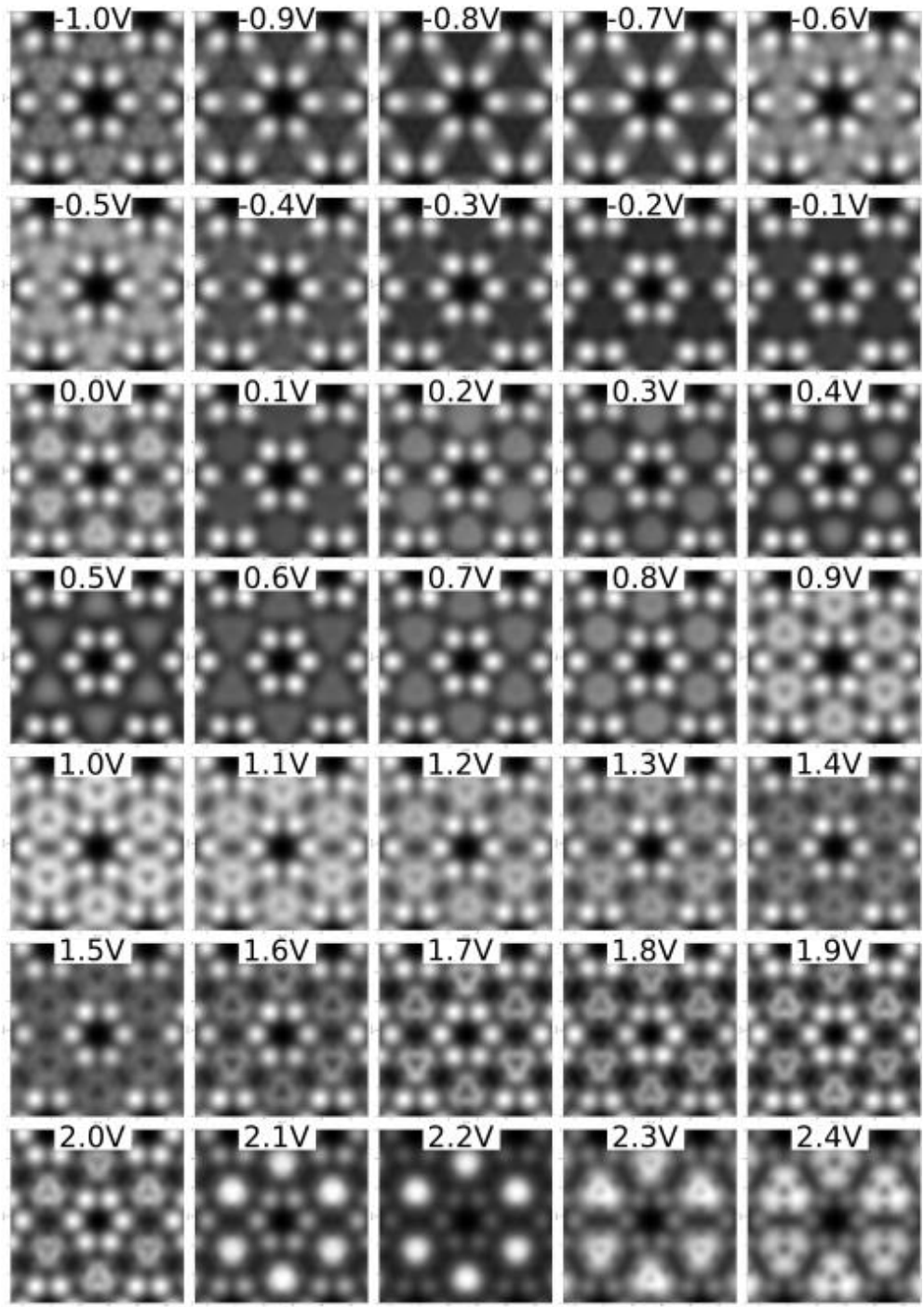

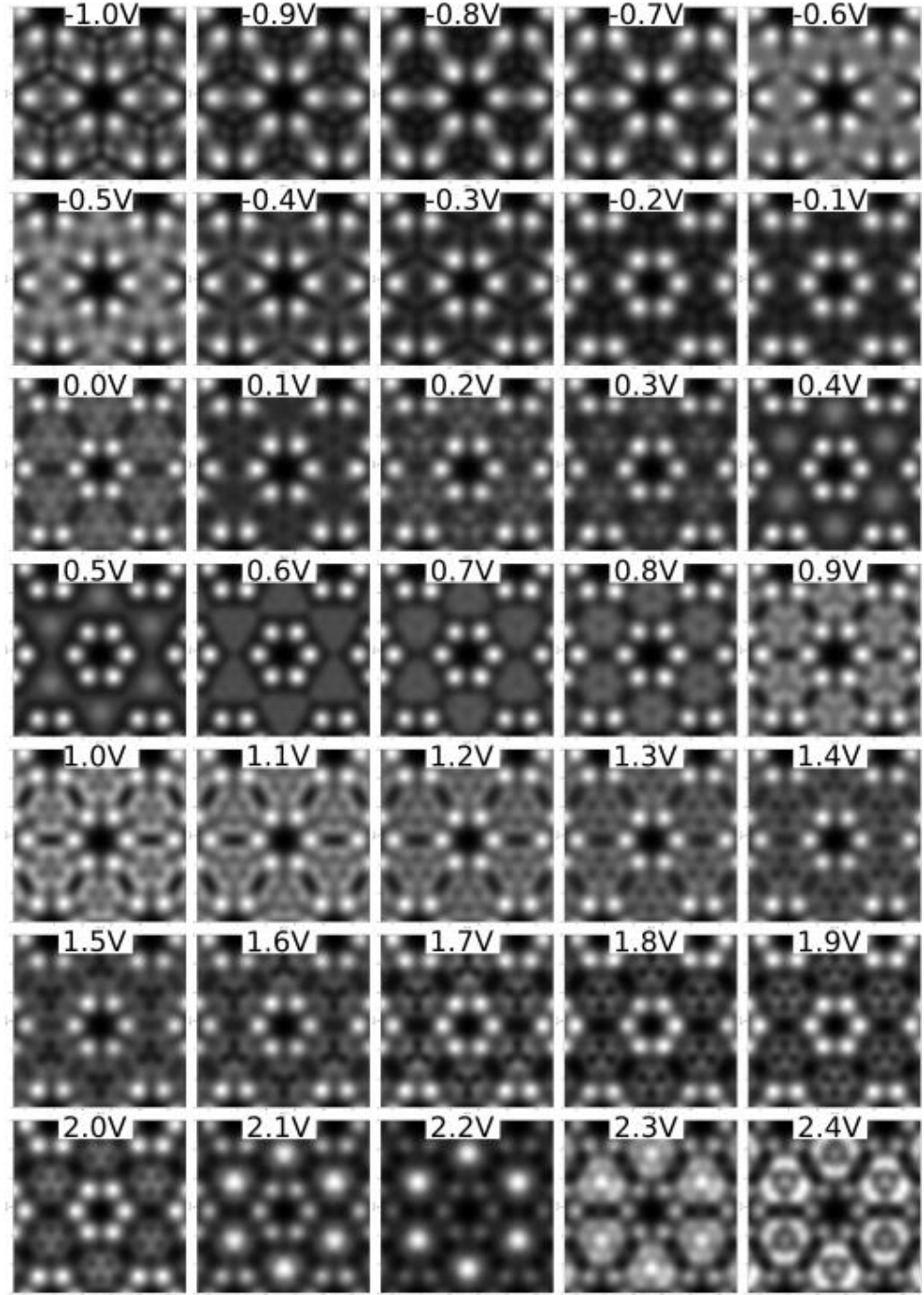

**Fig. S8.** A series of d*I*/d*V* constant height simulations obtained with mixed *s* and *p*-wave tips in order to simulate the CO tip in the experiment. The images were obtained through a linear combination of *p* ($p_x$ and $p_y$) and s orbitals, where we considered the portions shown in the top of the images.

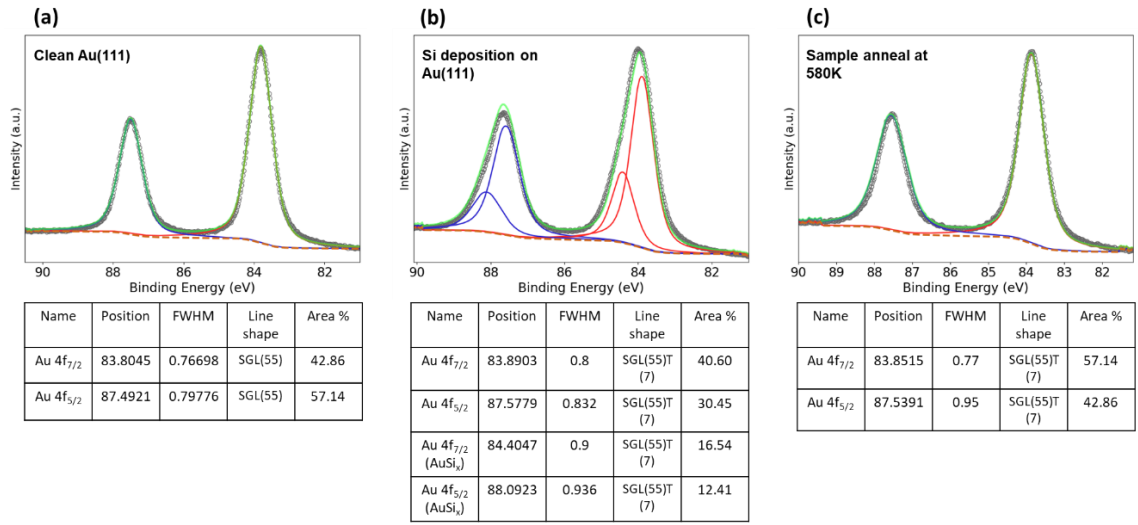

**Fig. S9**. High-resolution XPS Au 4f spectra taken after each experimental step. (**A**) Well-defined Au 4f spectra with doublet peaks separated by 3.69 eV. (**B**) After Si deposited on Au substrate, the Au 4f becomes broaden significantly. Two sets of doublets are following the same peak separation as well as the intensity ratio. (**C**) When the Si-COF is formed upon annealing at 580K, the Au 4f transformed back to the doublet peaks that are quite comparable with the clean surface situation.

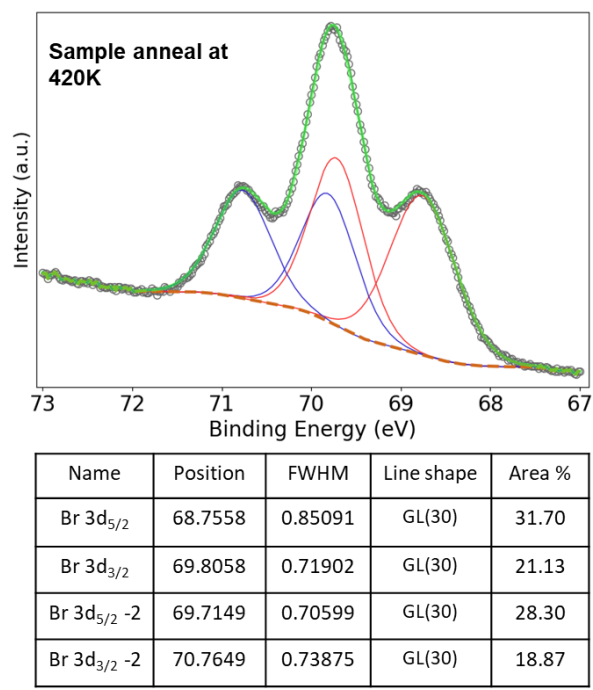

| Name | Position | FWHM | Line shape | Area % |
|---|---|---|---|---|
| Br $3d_{5/2}$ | 68.7558 | 0.85091 | GL(30) | 31.70 |
| Br $3d_{3/2}$ | 69.8058 | 0.71902 | GL(30) | 21.13 |
| Br $3d_{5/2}$ -2 | 69.7149 | 0.70599 | GL(30) | 28.30 |
| Br $3d_{3/2}$ -2 | 70.7649 | 0.73875 | GL(30) | 18.87 |

**Fig. S10.** High-resolution XPS Br 3d spectra at Si-COF formation after depositing HBTP precursor molecule at 420K. The doublet sets are best fitted with a separation of 1.05 eV and an intensity ratio of 0.67, and color coded as red and blue sets. Two sets of the doublet indicated two major types of Br-species at the intermediate stage, which possibly corresponds to disassociated $Br_2$ on surface and Si-Br bonded species.

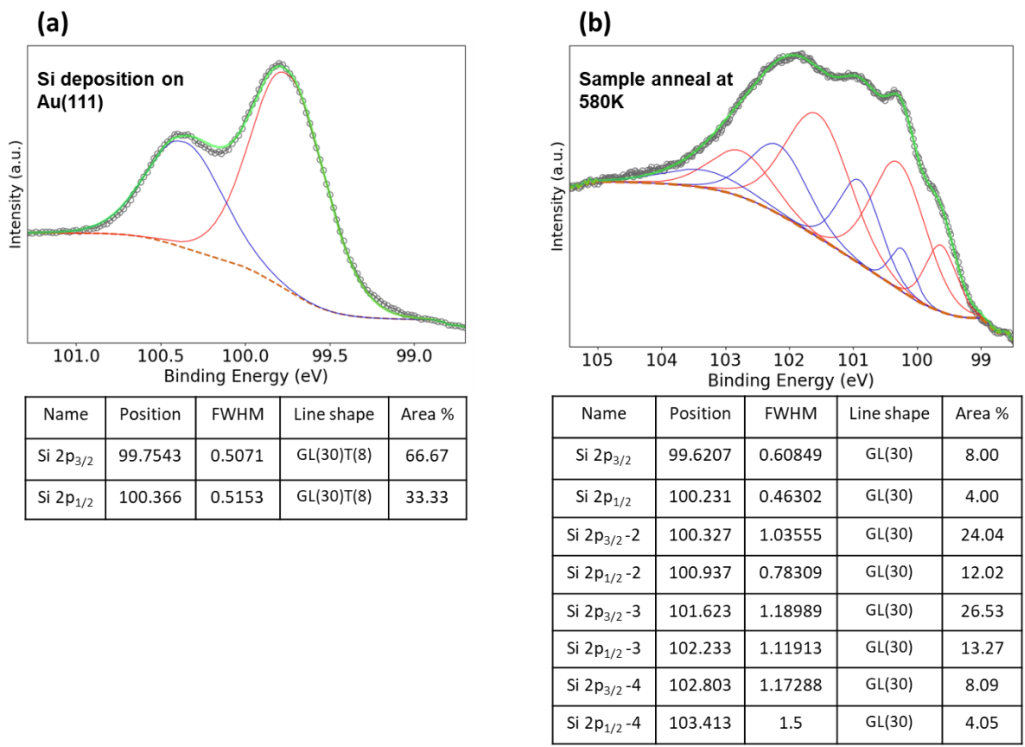

**Fig. S11.** High-resolution XPS Si 2p spectra after each experimental step. (**A**) The as-deposited Si on Au substrate is showing a well-defined doublet peaks with a slight shift to higher binding energy compare to that of bulk silicon peak positions. We attribute this with the formation of AuSi$_x$, as supported by Au 4f spectra at such experimental step. (**B**) Upon formation of Si-incorporated COF structure, the Si 2p became complicated convolutes shape that best fits with a minimum of 4 sets of doublets components, where predominantly two relatively large areas indicating a majority of two types of Si-species existence.

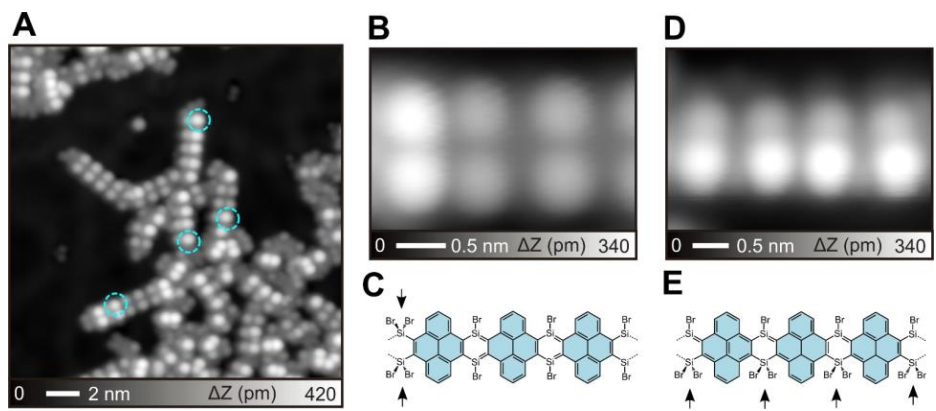

**Fig. S12.** (**A**) STM topography of Si-Cove GNRs on Au(111) after heating at 420 K. There are two types of bright dots. The brighter dots are indicated by circles. (**B**) Closeup view of one Si-Cove GNR with brighter dots, which are identified as two Br atoms bonded with one Si at the edge. The corresponding chemical structure was shown in (**C**). (**D**) Closeup view of another Si-Cove GNR with brighter dots. The corresponding chemical structure was shown in (**E**). Measurement parameters: $V = 100$ mV and $I = 5$ pA in (**A**). $V = 200$ mV and $I = 5$ pA in (**B,C**).

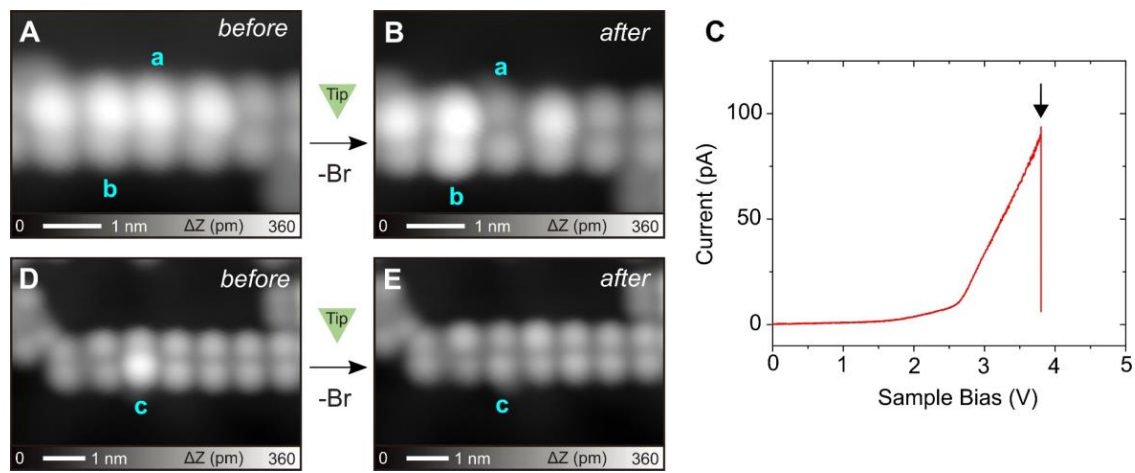

**Fig. S13.** Tip-induced manipulation on Si-Cover GNRs (**A**) STM topography of one Si-Cove GNR with four brighter dots on Au(111). Tip was put above the site of brighter dot **a** for manipulation. (**B**) STM image of Si-Cove GNR after sweeping the sample bias. The dot **a** has become darker due to removal of one Br, which has jumped and bonded with the Si atom at site **b**. The corresponding I-V curve was recorded in (**C**). At $V = 3.8$ V and $I = 90$ pA, the tunneling current jumped down, indicated that one Br atom has gone. STM topographies of another event of removing Br with tip manipulation: (**D**) before, (**E**) after. Measurement parameters: $V = 200$ mV and $I = 10$ pA in (**A**,**B**). $V = 200$ mV and $I = 20$ pA in (**D**,**E**).

**References**

(1) Zhao, D.; Wu, Q.; Cai, Z.; Zheng, T.; Chen, W.; Lu, J.; Yu, L. *Chem. Mater.* **2016**, *28*, 1139-1146.

(2) Bartels, L.; Meyer, G.; Rieder, K.-H. *Appl. Phys. Lett.* **1997**, *71*, 213−215.

(3) Watson, R. E.; Hudis, J.; Perlman, M. L. *Phys. Rev. B* **1971**, *4*, 4139–4144.

(4) Jankovský, O.; Šimek, P.; Klimová, K.; Sedmidubský, D.; Matejková, S.; Pumerac, M.; Sofer, Z. *Nanoscale* **2014**, *6*, 6065–6074.

(5) Smykalla, L.; Shukrynau, P.; Korb, M.; Lang, H.; Hietschold, M. *Nanoscale* **2015**, *7*, 4234–4241.

(6) Cardenas, L.; Gutzler, R.; Lipton-Duffin, J.; Fu, C.; Brusso, J. L.; Dinca, L. E.; Vondráček, M.; Fagot-Revurat, Y.; Malterre, D.; Rosei, F.; Perepichka, D. F. *Chem. Sci.* **2013**, *4*, 3263–3268.

(7) Eichhorn, J.; Strunskus, T.; Rastgoo-Lahrood, A.; Samanta, D.; Schmittele, M.; Lackinger, M. *Chem. Commun.* **2014**, *50*, 7680–7682.

(8) Kresse, G.; Furthmüller, J. *Comp. Mat. Sci.* **1996**, *6*, 15−50.

(9) Kresse, G.; Furthmüller, J. *Phys. Rev. B* **1996**, *54*, 11169−11186.

(10) Grimme, S.; Antony, J.; Ehrlich, S.; Krieg, S. *J. Chem. Phys.* **2010**, *132*, 154104.

(11) Grimme, S.; Ehrlich, S.; Goerigk, L. *J. Comp. Chem.* **2011**, *32*, 1456.

(12) Blochl, P. E. *Phys. Rev. B* **1994**, *50*, 17953−17979.

(13) Humphrey, W.; Dalke, A.; Schulten, K. *J. Mol. Graphics* **1996**, 14, 33−38.

(14) Otero-de-la-Roza, A.; Blanco, M. A.; Pendás, A. M.; Luaña, V. *Comput. Phys. Commun.* **2009**, *180*, 157–166.

(15) Otero-de-la-Roza, A.; Johnson, E. R.; Luaña, V. *Comput. Phys. Commun.* **2014**, *185*, 1007-1018.

(16) Tersoff, J.; Hamann, D. R. *Phys. Rev. B* **1985**, *31*, 805−813.

(17) Blum, V.; Gehrke, R.; Hanke, F.; Havu, P.; Havu, V.; Ren, X.; Reuter, K.; Scheffler, M. *Comput. Phys. Commun.* **2009**, *180*, 2175−2196.

(18) Perdew, J. P.; Burke, K.; Ernzerhof, M. *Phys. Rev. Lett.* **1996**, *77*, 3865.

(19) Krejčí, O.; Hapala, P.; Ondráček, M.; Jelínek, P. *Phys. Rev. B* **2017**, *95*, 045407.

(20) de la Torre, B.; Švec, M.; Foti, G.; Krejčí, O.; Hapala, P.; Garcia-Lekue, A.; Frederiksen, T.; Zbořil, R.; Arnau, A.; Vázquez, H; Jelínek, P. *Phys. Rev. Lett.* **2017**, *119*,

166001.

(21) Gross, L.; Moll, N.; Mohn, F.; Curioni, A.; Meyer, G.; Hanke, F.; Persson, M. *Phys. Rev. Lett.* **2011**, *107*, 086101.

(22) Wang, V.; Xu, N.; Liu, J. C.; Tang, G.; Geng, W. T. *Comput. Phys. Commun.* **2021**, *267*, 108033.